\def\papertitle{Realistic gramophone noise synthesis using a diffusion model}
\def\paperauthorA{Eloi Moliner}
\def\paperauthorB{Vesa V\"alim\"aki}

\documentclass[twoside,a4paper]{article}
\usepackage{etoolbox}
\usepackage{dafx_22a}
\usepackage{booktabs}
\usepackage{amsmath,amssymb,amsfonts,amsthm}
\usepackage{euscript}
\usepackage[T1]{fontenc}
\usepackage[utf8]{inputenc}
\usepackage{nimbusserif}
\usepackage{ifpdf}
\usepackage[english]{babel}
\usepackage{caption}
\usepackage{color}

\usepackage{subcaption}
\usepackage{algorithm}
\usepackage{algpseudocode}

\input glyphtounicode
\ninept

\newcounter{numauth}
\newcounter{listcnt}
\newcommand\authcnt[1]{\ifdefined#1 \stepcounter{numauth} \fi}

\newcommand\addauth[1]{
\ifdefined#1 
\stepcounter{listcnt}
\ifnum \value{listcnt}<\value{numauth}
\appto\authorslist{, #1}
\else
\appto\authorslist{~and~#1}
\fi
\fi}
\authcnt{\paperauthorB}
\authcnt{\paperauthorC}
\authcnt{\paperauthorD}
\authcnt{\paperauthorE}
\authcnt{\paperauthorF}
\authcnt{\paperauthorG}
\authcnt{\paperauthorH}
\authcnt{\paperauthorI}
\authcnt{\paperauthorJ}
\def\authorslist{\paperauthorA}
\addauth{\paperauthorB}
\addauth{\paperauthorC}
\addauth{\paperauthorD}
\addauth{\paperauthorE}
\addauth{\paperauthorF}
\addauth{\paperauthorG}
\addauth{\paperauthorH}
\addauth{\paperauthorI}
\addauth{\paperauthorJ}

\usepackage{times}

\newif\ifpdf
\ifx\pdfoutput\relax
\else
   \ifcase\pdfoutput
      \pdffalse
   \else
      \pdftrue
\fi

\ifpdf 
  \usepackage[pdftex,
    pdftitle={\papertitle},
    pdfauthor={\authorslist},
    pdfsubject={Proceedings of the 25th International Conference on Digital Audio Effects (DAFx20in22)},
    colorlinks=false, 
    bookmarksnumbered, 
    pdfstartview=XYZ 
  ]{hyperref}
  \usepackage[pdftex]{graphicx}
\else 
  \usepackage[dvips]{epsfig,graphicx}
  \usepackage[dvips,
    pdftitle={\papertitle},
    pdfauthor={\authorslist},
    pdfsubject={Proceedings of the 25th International Conference on Digital Audio Effects (DAFx20in22)},
    colorlinks=false, 
    bookmarksnumbered, 
    pdfstartview=XYZ 
  ]{hyperref}
\fi
\usepackage[hypcap=true]{caption}
\title{\papertitle}

\affiliation{
\paperauthorA\ and  \paperauthorB\, \thanks{\vspace{-3mm}}}
{\href{https://www.aalto.fi/en/aalto-acoustics-lab}{Acoustics Lab}, Dept. of Signal Processing and Acoustics \\ Aalto University \\ Espoo, Finland\\
{\tt \href{mailto:eloi.moliner@aalto.fi}{eloi.moliner@aalto.fi}}
}

\begin{document}
\ifpdf 
  \DeclareGraphicsExtensions{.png,.jpg,.pdf}
\else  
  \DeclareGraphicsExtensions{.eps}
\fi


\maketitle

\begin{abstract}
This paper introduces a novel data-driven strategy for synthesizing gramophone noise audio textures. A diffusion probabilistic model is applied to generate highly realistic quasiperiodic noises. 
The proposed model is designed to generate samples of length equal to one disk revolution, but a method to generate plausible periodic variations between revolutions is also proposed.
A guided approach is also applied as a conditioning method, where an audio signal generated with manually-tuned signal processing is refined via reverse diffusion to improve realism.
The method has been evaluated in a subjective listening test, in which the participants were often unable to recognize the synthesized signals from the real ones. The synthetic noises produced with the best proposed unconditional method are statistically indistinguishable from real noise recordings. This work shows the potential of diffusion models for highly realistic audio synthesis tasks.
\end{abstract}

\section{Introduction}
\label{sec:intro}
 \sloppy
The quality of audio recording has improved greatly over the last century. Far away are those old gramophone recordings that our ancestors used to listen to. During the gramophone era, recorded music sounded highly distorted, bandlimited, and, before anything else, notoriously noisy. Although the sound quality of gramophone recordings is very poor compared to modern standards, there is a certain interest in emulating these characteristic sounds of the past. 
Gramophone recordings comprise a rich amount of heterogeneous noises that are appealing for creative uses.
The idea of processing modern audio files to appear aged by imitating historical disturbances is named in the literature as digital audio antiquing \cite{vlimki2008digital}.  This paper focuses on the particular task of imitating all kinds of additive noise that appear in gramophone recordings.

A simple way of applying realistic gramophone noises to a piece of music is to directly use noise extracts from real recordings. The idea would consist of extracting a noise sample, looping it, and directly adding it to the music signal. However, this method would require a time-consuming search to find the desired noise characteristics, in addition to having access to a collection of digitized gramophone recordings. Moreover, if the extracted noise sample is not long enough, the looping effect could also be perceived by a listener. As a consequence, there is a need for more practical and versatile methods to generate such sounds.



The research on audio antiquing started in Aalto Acoustics Lab almost 20 years ago, when a science museum had interest in it. The original idea was to show how music listening had changed over 100 years by applying simulated degradations of music media to the same music piece. This previous study \cite{vlimki2008digital} consisted of using digital signal processing techniques to model the degradations in historical recordings, including the most relevant additive noises in gramophones, such as hisses, clicks, and thumps. Although this method is accurate in simulating some of the degradations, it is unsuccessful in synthesizing realistic clicks and scratches\footnote{\href{http://research.spa.aalto.fi/publications/papers/antiquing/}{http://research.spa.aalto.fi/publications/papers/antiquing/}}. Another closely related work is the freely available plugin \textit{iZotope Vinyl} \cite{izotopevinyl}, which simulates the degradations of LP recordings, but the underlying details of the method are not publicly available.

However, these previous approaches do not account for the wide and complex distribution of different noises that may appear in gramophone recordings. This is why a data-driven generative model may be advantageous. Generative models learn to produce new data instances by capturing the data distribution. Diffusion probabilistic models \cite{ho2020denoising}, often referred to as diffusion models, are a new class of generative models, some examples of recent applications include producing high-quality images  \cite{dhariwal2021diffusion}, speech \cite{kong2020diffwave, chen2020wavegrad}, drum sounds \cite{rouard2021crash}, or symbolic music \cite{mittal2021symbolic}. Here, we adopt these methods to generate realistic gramophone noises with a data-driven approach.

The rest of this paper is organised as follows. Sec.~\ref{sec:noises} introduces the characteristics of the gramophone noises that we are interested in synthesizing. Sec.~\ref{sec:diffusion} introduces diffusion models. In Sec.~\ref{sec:methods}, we describe the presented methods. Sec.~\ref{sec:results} shows our evaluation results and, finally, Sec.~\ref{sec:conclusion} concludes.

\section{Gramophone Noises}\label{sec:noises}

Gramophone recordings contain a composite of noises produced by many different sources. This section summarizes the main characteristics of these noises and some methods used to model them.

\subsection{Hiss}

Hiss is the term used to refer to background noise. In gramophone recordings, hisses are often very noticeable and may come from a mixture of different sources, such as electrical circuit noise, irregularities in the storage medium, or ambient noise from the recording environment \cite{godsill_digital_1998}. Hisses are present throughout the duration of the recording and hence are referred to as global degradations \cite{vlimki2008digital}. However, they cannot be considered stationary because the sources that produce them are usually time-varying. A particular example is the ``swishing'' effect related to periodic patterns that often appear when, for instance, one part of the disk is damaged more than the other parts. 

Every single gramophone recording ends up containing a unique hiss sound, due to the wide variety of stochastic sources that produce them. While it is possible to model one particular kind of hiss using manually-tuned classical signal processing \cite{vlimki2008digital}, it is not possible in practice to capture the wide distribution of hiss noises with this approach.

\subsection{Clicks}

Clicks are one of the most recognizable degradations in old recordings. They are localized impulsive disturbances mainly caused by dust particles or small scratches on the media surface. Typically, individual clicks have a short duration, and their durations range from less than 20 $\mu$s to 4\,ms \cite{rayner1991detection}. The amplitude of the clicks can also vary greatly within the same recorded extract \cite{godsill_digital_1998}. In gramophone recordings, the rate of click occurrence is remarkably dense, about 2,000 clicks per second. In these cases, only the strongest clicks can be heard as individual elements. The largest remaining clicks are perceived as a cumulus of click bursts, often named as ``crackle''. 

In audio restoration, clicks are often modeled as additive disturbances \cite{godsill_digital_1998, rund2021evaluation}. With some prior knowledge on the statistics of clicks---e.g., duration, amplitude, and frequency of occurrence---it is possible to synthesize them using random numbers \cite{vlimki2008digital}. However, using this methodology, the synthesized clicks appear to be  static, but in reality their timbre is constantly changing.

%
%
 

\subsection{Low-frequency pulses}
Low-frequency pulses, often called ``thumps'', are a recognizable type of degradation associated with severely damaged old recordings  \cite{esquef2003an}. They are produced by strong discontinuities in the media, such as deep scratches or when two broken parts of the disk are fixed with glue. This degradation can generally be described by a short and strong discontinuity followed by a low-frequency tail \cite{esquef2003an}. The initial discontinuity occurs when the stylus passes through the physical discontinuity. It usually lasts less than 2\,ms and behaves as high-variance noise added to the original signal. The tail consists of damped oscillations of decaying frequency caused by the impulse response of the arm, normally lasting longer than 50\,ms. 

Following Esquef \emph{et al}.~\cite{esquef2003an}, the tail part of a thump can be modeled in the time domain as:
\begin{equation}\label{thumpmodel}
    s_\text{tail}(n)=A_\text{tail} e^{-n / f_\text{s} \tau_\text{e}} \sin \left( 
    2\pi n \frac{f_n}{f_\text{s}} - \frac{\pi}{4}
    \right),
\end{equation}
where $f_\text{s}$ is the sampling frequency, $A_{\text{tail}}$ is the tail amplitude, $\tau_\text{e}$ is a time constant associated with the envelope decay and the variation in the oscillation frequency $f_n$ can be modeled in the following way: 
\begin{equation}
    f_n=(f_\text{max}- f_\text{min}) e^{-n/f_\text{s} \tau_\text{f}} + f_\text{min}, 
\end{equation}
where $f_\text{max}$ and $f_\text{min}$ are the maximum and minimum oscillation frequencies and $\tau_\text{f}$ is another time constant that describes the frequency-decay rate. The tail parameters may vary depending on the severity of the scratch.

\subsection{Other noises}
Apart from the aforementioned disturbances, there are other elements that make gramophone noises more diverse and interesting.  Some digitized recordings present the characteristic ``humming'' sound caused by high-voltage electricity at the alternating current frequency of 50\,Hz or 60\,Hz. This interference signal normally contains higher harmonics that are easily perceptible. Another characteristic example are the rumble noises of low-frequency content caused by vibrations of the turntable.


\section{Diffusion Models}\label{sec:diffusion}


Diffusion models are a new class of generative models comprised of a forward diffusion process and a reverse diffusion process \cite{ho2020denoising}. During the forward diffusion, the data is progressively transformed to a tractable prior distribution, usually a standard Gaussian. The reverse diffusion process is parameterized with a neural network and has the goal of reversing the diffusion process by iteratively denoising the diffused data. Once the model has been trained, the reverse diffusion process defines a mapping between the chosen prior distribution and the training data. In this section, we summarize the details using the notation from \cite{kingma2021variational}.

\subsection{Forward diffusion process}

We start by defining a diffusion process $\{\mathbf{z}_\tau\}_{\tau=0}^1$, indexed by a continuous variable $\tau \in [0,1]$\footnote{We use the index variable $\tau$ instead of $t$ to avoid confusion with the time variable in audio signals.}. The goal of the diffusion process is to progressively transform data samples $\mathbf{x} \sim p_\text{data}$ to a tractable prior distribution $\mathbf{z}_1 \sim  p_\text{prior}$. The distribution of the latent variables $\mathbf{z}_\tau$ conditioned on $\mathbf{x}$, for any diffusion step $\tau$, is given by:
\begin{equation}\label{closedform}
   q(\mathbf{z}_\tau|\mathbf{x})=\mathcal{N}(\alpha_\tau \mathbf{x}, \sigma_\tau^2 \mathbf{I}),
\end{equation}
where $\alpha_\tau$ is a mean coefficient and $\sigma_\tau^2$ is the noise variance, which are both continuous positive functions in the range $[0,1]$.  We focus only on the \textit{variance-preserving} diffusion \cite{sohl2015deep, ho2020denoising}, where  the two functions are directly related as $\alpha_\tau=\sqrt{1-\sigma^2_\tau}$. In this particular case, $\alpha_1 \approx 0$, and $\sigma_1 \approx 1$ when $\tau=1$. Thus, the prior distribution $p_\text{prior}$ converges to a standard normal $\mathcal{N}(\mathbf{0},\mathbf{I})$. We emphasize the notion that $\mathbf{z}_\tau$ gets increasingly noisy as we go forward in the diffusion step $\tau$. The noise variance $\sigma^2_\tau$ usually has a fixed predefined schedule, which is required to be smooth and monotonically increasing. In particular, we apply the schedule:
\begin{equation}
    \sigma_\tau=\left(1-\cos(\pi \tau)\right)/2,
\end{equation}
which is the same as in \cite{rouard2021crash}, and shares similarities with the one proposed in \cite{nichol2021improved}. 

In practice, the diffusion process is discretised in a finite number of diffusion steps $T$ that behave as a Markov chain \cite{ho2020denoising}. As defined in \cite{kingma2021variational}, given $s=\tau-\frac{1}{T}$ and $0<s<\tau<1$, the conditional forward distribution $q(\mathbf{z}_\tau| \mathbf{z}_s)$ is given by:
\begin{equation}\label{forwardconf}
   q(\mathbf{z}_\tau| \mathbf{z}_s)=\mathcal{N}\left(\alpha_{\tau|s}\mathbf{z}_s, \sigma_{\tau|s}^2 \mathbf{I}\right),
\end{equation}
where
\begin{equation}
    \alpha_{\tau|s}=\alpha_\tau / \alpha_s,
\end{equation}
and  
\begin{equation}
    \sigma_{\tau|s}^2=\sigma_\tau^2 -\alpha_{\tau|s}^2 \sigma_s^2.
\end{equation}
After specifying the forward diffusion process, we are now ready to define the reversed version of it in the following section. 

\subsection{Reverse diffusion process}
\begin{figure}[t]
    \centering
    \includegraphics[trim={40 60 20 3}, clip]{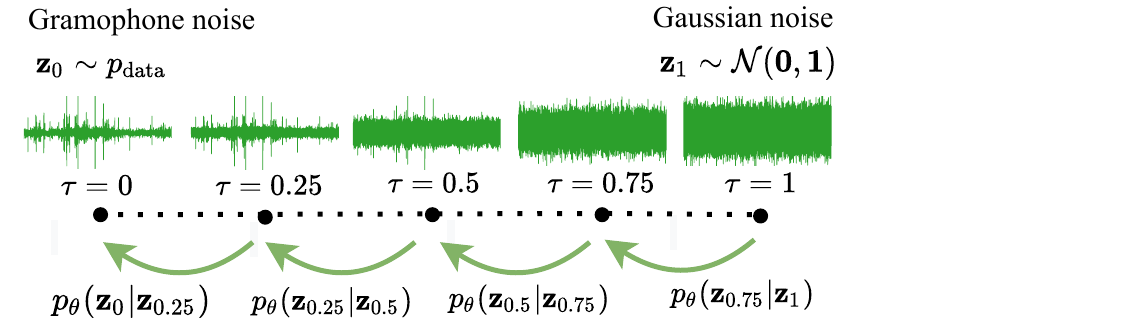}
    \caption{{\it Example of a reverse diffusion process discretized in $T=4$ steps.}}
    \label{fig:reversediff}
\end{figure}

Given Eq.~\eqref{closedform} and Eq.~\eqref{forwardconf}, applying the Bayes rule, it can be proven that, for any $0<s<\tau<1$, the reverse conditional distribution $q(\mathbf{z}_s|\mathbf{z}_\tau, \mathbf{x})$ is also Gaussian and can be defined as:
 \begin{equation}
   q(\mathbf{z}_s| \mathbf{z}_\tau, \mathbf{x})=\mathcal{N}\left(
   \frac{\alpha_{\tau|s} \sigma_s^2}{\sigma_\tau^2}
   \mathbf{z}_t
   +\frac{\alpha_s \sigma_{\tau|s}^2}{\sigma_\tau^2}\mathbf{x}
   ,
 \frac{\sigma_{\tau|s}^2 \sigma_s^2}{\sigma_\tau^2}\mathbf{I}
   \right).
 \end{equation}
 However, since the input data $\mathbf{x}$ is not known during inference, the reverse conditional distribution is modeled as follows:
 \begin{equation}\label{reversestep}
     p_\theta(\mathbf{z}_s| \mathbf{z}_\tau)=q(\mathbf{z}_s| \mathbf{z}_\tau, \mathbf{x}= \mathbf{\hat{x}}_\theta(\mathbf{z}_\tau, \sigma_\tau) )
 \end{equation}
 where $\mathbf{\hat{x}}_\theta(\mathbf{z}_\tau, \sigma_\tau)$ is the output of a denoising model. As it is commonly used \cite{ho2020denoising, dhariwal2021diffusion, kingma2021variational}, we apply a change of variables to utilize a noise estimation model instead by:
 \begin{equation}
 \hat{\mathbf{x}}_\theta(\mathbf{z}_\tau, \sigma_\tau)=
      (\mathbf{z}_\tau -\sigma_\tau \hat{\boldsymbol{\epsilon}}_\theta(\mathbf{z}_\tau, \sigma_\tau)
     )/\alpha_\tau,
 \end{equation}
where $ \hat{\boldsymbol{\epsilon}}_\theta(\mathbf{x}_\tau,\sigma_\tau)$ is the output of a deep neural network. 

During training, the noisy signal $\mathbf{z}_\tau$ is generated from the training data $\mathbf{x} \sim p_\text{data} $ following Eq.~\eqref{closedform} as:
\begin{equation}\label{perturb}
\mathbf{z}_\tau= \alpha_\tau\mathbf{x}+ \sigma_\tau \boldsymbol\epsilon,
\end{equation}
and the neural network $ \hat{\boldsymbol{\epsilon}}_\theta(\mathbf{x}_\tau,\sigma_\tau)$ is optimized to estimate the added noise  $\boldsymbol\epsilon \sim \mathcal{N}(\mathbf{0}, \mathbf{I})$. Since, given the input data $\mathbf{x}$, we can easily produce the latent variable $\mathbf{z}_\tau $ at any diffusion step $\tau$ using Eq.~\eqref{perturb}, the training can be conveniently performed at random uniformly-sampled diffusion steps. The learning objective can be defined as a weighted L2 loss:
\begin{equation}
L=\mathbb{E}_{\mathbf{x} \sim p_\text{data}, \boldsymbol\epsilon \sim \mathcal{N}(\mathbf{0},\mathbf{I}),  \tau \sim  \mathcal{U}([0,1])} [\lambda(\tau)
\Vert \boldsymbol\epsilon - \hat{\boldsymbol{\epsilon}}_\theta(\mathbf{z}_\tau, \sigma_\tau) \rVert_{2}
],
\end{equation}
where $\lambda(\tau)$ is a positive weighting function. We use a constant weighting $\lambda(\tau)=1$, as proven to be effective in \cite{ho2020denoising} and \cite{song2020score}. Following \cite{chen2020wavegrad}, the neural network is conditioned on the noise level $\sigma_\tau$, instead of directly on the diffusion step $\tau$. This implementation choice allows for flexibility during sampling, as the number of discretization steps $T$ and the noise variance schedule can then be modified without the need of retraining the model. The training procedure utilised in this work is also summarized in Alg. \ref{alg:cap}.
\begin{algorithm}[t]
\caption{Training}\label{alg:cap}
\begin{algorithmic}
\Repeat
\State Sample $\mathbf{x} \sim p_{\text{data}}$, $\boldsymbol\epsilon \sim \mathcal{N}(\mathbf{0},\mathbf{I})$ and $\tau \sim  \mathcal{U}([0,1])$
\State $\mathbf{z}_\tau \gets \alpha_\tau\mathbf{x} + \sigma_\tau \boldsymbol\epsilon$ 
\State Take gradient step on $\nabla_{\theta} \Vert \boldsymbol\epsilon - \hat{\boldsymbol\epsilon}_{\theta}(\mathbf{z}_\tau, \sigma_\tau) \rVert_{2}$
\Until{convergence}
\end{algorithmic}
\end{algorithm}

Once the training has converged, it is possible to reverse the forward diffusion process and obtain samples from the training data distribution $\mathbf{x} \sim p_\text{data}$ by starting from the prior $\mathbf{z_1} \sim p_\text{prior}$, which in our case is Gaussian noise. To do so, we discretize the diffusion process in $T$ steps and iterate over Eq.~\eqref{reversestep}, which can be applied as:
\begin{equation}\label{samplingstep}
    \mathbf{z}_s= f(\tau,s) \mathbf{z}_\tau -g(\tau,s) \hat{\boldsymbol\epsilon}_{\theta}(\mathbf{z}_{\tau}, \sigma_{\tau}) + h(\tau,s)\boldsymbol\epsilon,
\end{equation}
where $\boldsymbol\epsilon \sim \mathcal{N}(\mathbf{0}, \mathbf{I})$ is Gaussian noise and the functions $f(\tau,s)$, $g(\tau,s)$ and $h(\tau,s)$ are, respectively, defined as
\begin{equation}
    f(\tau,s)=\frac{1}{\alpha_{\tau|s}}, \;\;
g(\tau,s)=\frac{\sigma_\tau^2}{\alpha_{\tau|s}}-\frac{\alpha_{\tau|s}\sigma_s^2}{\sigma_\tau},
\end{equation}
and
\begin{equation}
   h(\tau,s)=\frac{\sigma_{\tau|s}^2 \sigma_s^2}{\sigma_\tau^2}.
\end{equation}
The number of discretization steps $T$ is a hyperparameter that defines a trade-off between inference quality and computational complexity of the inference.
A graphical example of a reverse diffusion process is represented in Fig.~\ref{fig:reversediff}
The practical implementation of the sampling algorithm is also summarized in Alg. \ref{alg:sampling}.

\begin{algorithm}[t]
\caption{Unconditional sampling}\label{alg:sampling}
\begin{algorithmic}
\Require $T$ (num. steps)
\State Sample  $\mathbf{z_1} \sim \mathcal{N}(\mathbf{0},\mathbf{I})$
\For{$i=T-1, \dots, 0$}
\State $s \gets \frac{i}{T}$, $\tau \gets \frac{i+1}{T}$
\State $\mathbf{z}_{s} \gets f(\tau,s) \mathbf{z}_{\tau} - g(\tau,s)\hat{\boldsymbol\epsilon}_{\theta}(\mathbf{z}_{\tau}, \sigma_{\tau})  $ 
\If{$s>0$} 
\State Sample $\boldsymbol\epsilon \sim \mathcal{N}(\mathbf{0},\mathbf{I})$
\State $\mathbf{z}_{s} \gets \mathbf{z}_{s}+ h(\tau,s)\boldsymbol\epsilon$
\EndIf
\EndFor
\end{algorithmic}
\end{algorithm}

\section{Methods}\label{sec:methods}

\subsection{Diffusion model for gramophone noise synthesis}\label{applying}

We apply the diffusion model presented in Sec.~\ref{sec:diffusion} for the task of synthesizing gramophone recording noise. To do so, we use the Gramophone Recording Noise Dataset, a collection of only-noise excerpts extracted from real gramophone recordings, which was previously used to train a denoising model \cite{Moliner2022}. Considering the periodic structure that gramophone recording noises present, we opt for defining the length of the time frames as the revolution period. Given that gramophone records were played at about 78 revolutions per minute, the frame length is defined as $L=60/78\approx0.77$\,s. Thus, at each training iteration, we extract random chunks of length $L$ from the dataset.

We noticed that diffusion models are very sensitive to energy differences in the data. A model trained with a wide variability of energy levels tends to produce samples at random volumes, which is not often a desired property in audio synthesis. To correct for this, we normalize the training noise segments by their root-mean-squared value. However, we decided to use the median instead of the standard mean to avoid the local disturbances, such as strong clicks or thumps, from affecting the energy estimation, given that the median is less influenced by outliers than the mean. The audio samples are normalized by:  
\begin{equation}\label{normalization}
    \mathbf{\bar{x}}=\frac{G}{\sqrt{b_\chi^2 + \text{median}(\mathbf{x}^2)}}\mathbf{x},
\end{equation}
where $\text{median}(\cdot)$ is the median operator, $b_{\chi}=1.4826$ is a constant that relates the median with the mean assuming a Gaussian distribution \cite{rousseeuw1993alternatives}, and $G=-10$\,dB is the applied constant gain.

\begin{figure}[t]
    \centering
    \includegraphics[scale=0.9,trim={8 5 90 1}, clip]{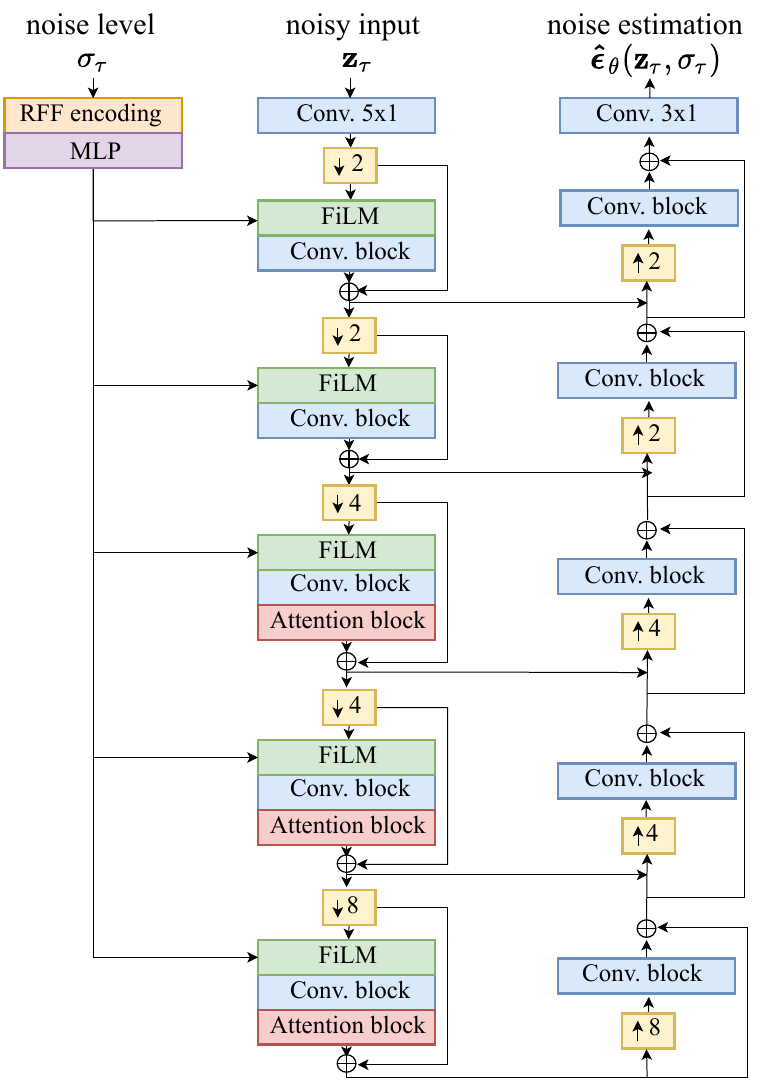}
    \caption{{\it Diagram of the noise estimation neural network $\boldsymbol{\hat\epsilon_\theta}(\mathbf{z_\tau}, \sigma_\tau)$, inspired by \cite{rouard2021crash}. The term ``RFF encoding'' refers to the noise level embedding based on Random Fourier Features \cite{tancik2020fourier}, ``MLP'' is a multi-layer perceptron, ``Conv. block'' is a stack of dilated convolutions, and ``Attention block'' refers to a multi-head self-attention layer \cite{vaswani2017attention}.}}
    \label{architecture}
\end{figure}

For the neural network for noise prediction $\epsilon_\theta(\mathbf{z}_\tau, \sigma_\tau)$, we use a similar architecture as Rouard and Hadjeres \cite{rouard2021crash}, which consists of a U-Net in the time domain represented in Fig.~\ref{architecture}. The noise level $\sigma_\tau$ is encoded using random Fourier features \cite{tancik2020fourier} and a multi-layer perceptron, whose output is applied as a conditioning signal using FiLM layers \cite{perez2018film}. Different to \cite{rouard2021crash}, we add multi-head self-attention layers \cite{vaswani2017attention} after the third, fourth, and fifth blocks. This architectural choice yields a global receptive field to the model and helps to capture slowly-varying textures within the revolution period. Circular padding is used in the convolutional layers as an inductive bias to fit the periodicity of the data.
We refer to \cite{rouard2021crash} and 
the source code\footnote{\href{https://github.com/eloimoliner/gramophone_noise_synth}{https://github.com/eloimoliner/gramophone\_noise\_synth} } for more details on the architecture.

 We use the Adam optimizer with a learning rate equal to \mbox{$2\cdot10^{-4}$}. The weights are smoothed with exponential moving average at a rate of 0.999. We train the diffusion model using Alg. \ref{alg:cap} for a total of 750k iterations using a batch size of 16. The training took 48 h to complete using a single NVIDIA A100 GPU in Triton, Aalto University's computing cluster.

\subsection{Guided synthesis}\label{datadriverefinement}
\begin{figure*}[t]
    \centering
    \includegraphics[trim={10 5 190 1}, clip]{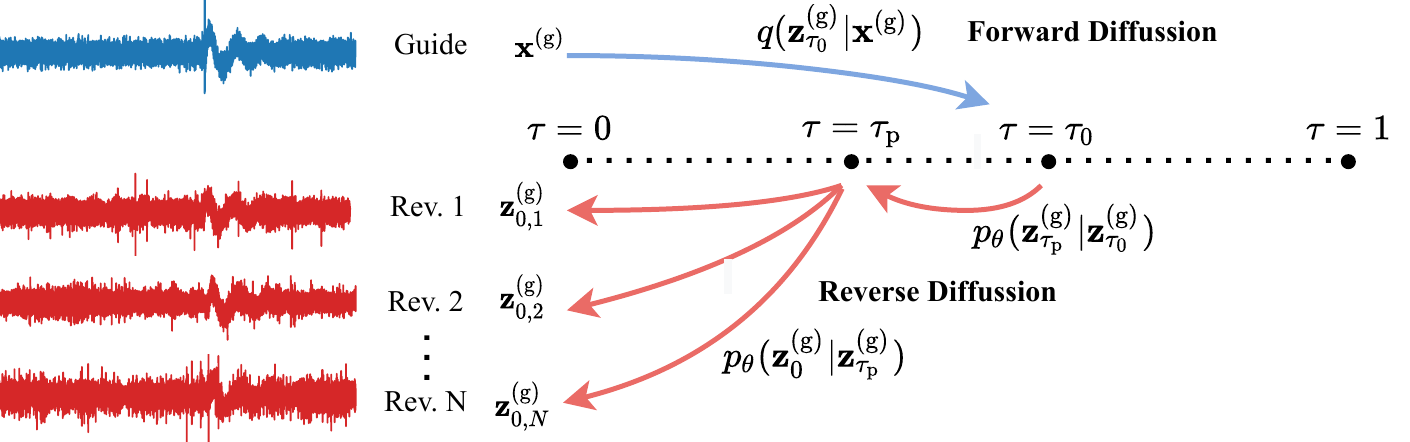}
    \caption{{\it Diagram of the guided synthesis method. The guided synthesis starts with a manually-designed reference signal (guide) that is iteratively refined via reverse diffusion, starting from a truncation step $\tau_0$, to generate a more realistic output. The reverse diffusion process is bifurcated at an intermediate step $\tau_\text{p}$ to output a set of revolutions, which are correlated to the guide but present notable variations. }}
    \label{guided_diffusion}
\end{figure*}

%
\begin{figure*}[t]
    \centering
    \includegraphics[scale=0.95, trim={0 7 0 0}, clip]{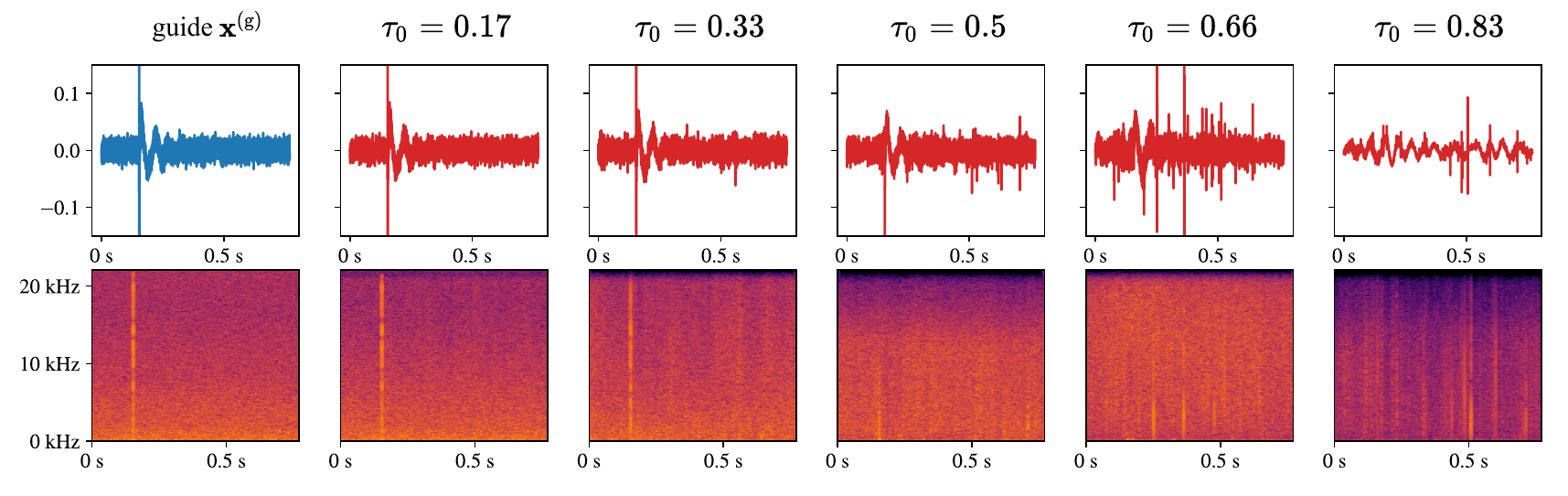}
    \caption{{\it Waveform and spectrogram representations of single period (0.77\,s) of gramophone noise produced using different values of $\tau_0$.}}
    \label{tau0}
\end{figure*}
\begin{figure}[t]
    \centering
         \begin{subfigure}{\linewidth}
    \includegraphics[scale=0.35]{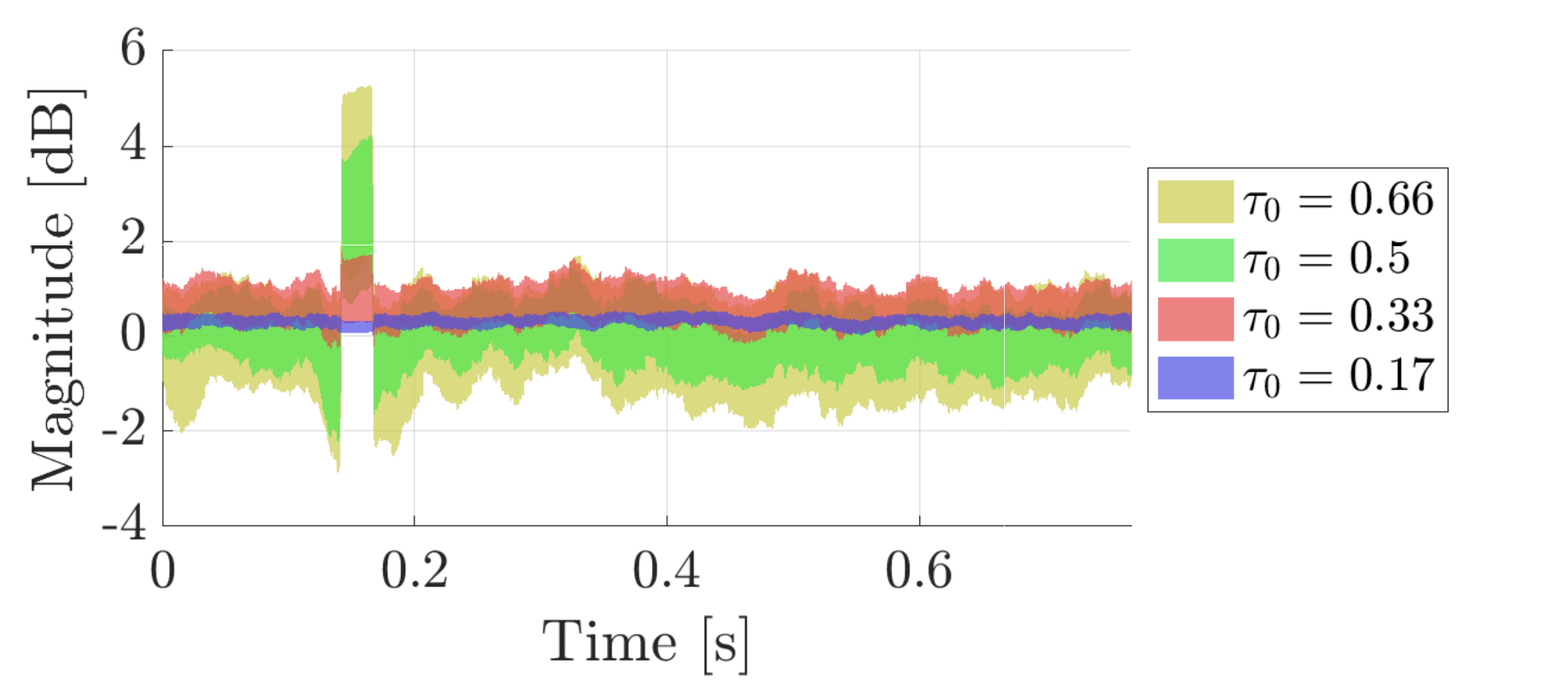}
    \caption{Variations in the temporal envelope for different values of $\tau_0$.}
    \label{stat_time}
        \end{subfigure}
     \begin{subfigure}{\linewidth}
    \includegraphics[scale=0.35]{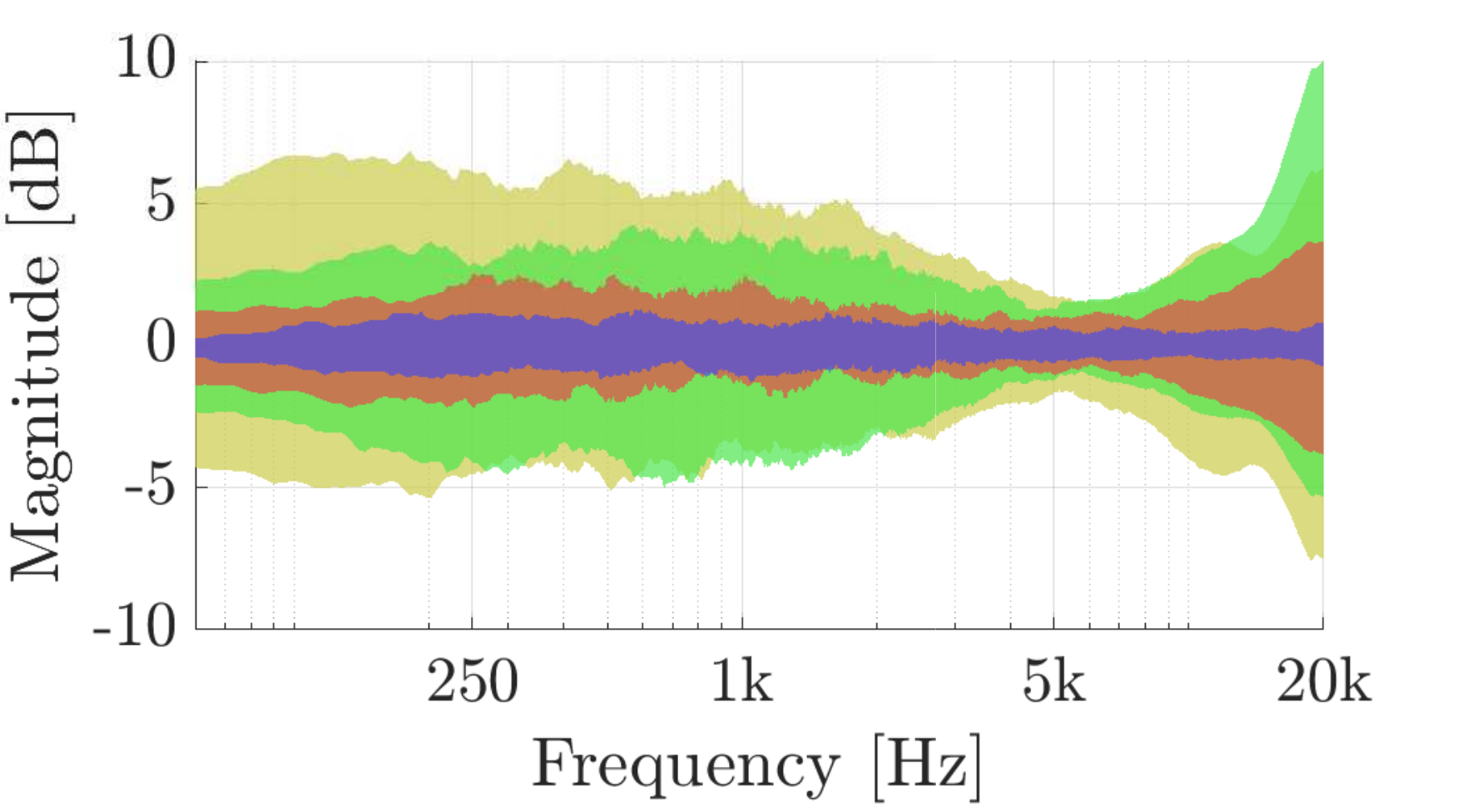}
    \caption{Variations in the spectral envelope for different values of $\tau_0$.}
         \label{stat_freq}
     \end{subfigure}
     \caption{{\it Analysis of the effect of the truncation step $\tau_0$ on the variations between a guide and a set of refined samples generated using guided synthesis. The standard deviations of the differences between (a) the temporal envelopes and (b) the Bark-smoothed spectral envelopes are plotted with different colors representing different values of $\tau_0$. }}
     \label{deviations_tau0}
\end{figure}
Once the model described in Sec.~\ref{applying} has been trained, we can use Alg. \ref{alg:sampling} to unconditionally sample noise frames $\mathbf{z}_0$. As demonstrated in Sec.~\ref{sec:results}, these generated samples can be highly realistic and have a wide range of variability. However, one could argue that such an unconditional synthesis model would provide no practical advantage against using real noise samples directly. In this section, we propose a method for conditioning the reverse diffusion process, which can be used for synthesizing realistic noises with a certain degree of control over their characteristics, allowing plausible variations. 

 In a recent work on guided image synthesis \cite{meng2021sdedit}, Meng \emph{et al}.~show how the reverse diffusion process can be used to refine a signal $\mathbf{x}^\text{(g)}$ referred to as the ``guide''. In our particular case, the guide is thought to be an audio segment of length $L$, i.e., one disk revolution, which serves as a draft for the noise texture $\mathbf{z}_0$ we are interested to generate. We can synthesize $\mathbf{x}^\text{(g)}$ using known parameterizable digital signal processing techniques \cite{vlimki2008digital}, as detailed in Sec.~\ref{sec:guides} which offer high interpretability but often lack realism.
 
 The general idea is to start the reverse diffusion process from an intermediate truncation step $\tau_0 \in [0,1]$, and the starting point is the perturbed guide:
 \begin{equation}
 \mathbf{z}^{(\text{g})}_{\tau_0} \sim \mathcal{N}( \alpha_{\tau_0}\mathbf{x}^\text{(g)},\sigma^{2}_{\tau_0}\mathbf{I}).
 \end{equation}
As we do for the unconditional sampling, starting from an intermediate diffusion step $\mathbf{z}^{(\text{g})}_{\tau_0}$, we can iterate over Eq.~\eqref{reversestep} until we obtain $\mathbf{z}^{(\text{g})}_0$, which would ideally be closer to the training data distribution. In summary, using guided sampling, we have defined a mapping between a simplistically synthesized noise texture $\mathbf{x}^{(\text{g})}$ to a refined version of it $\mathbf{z}^{(\text{g})}_{0}$ with similar characteristics, but more realistic sounding. The guided sampling procedure is also summarized in Alg. \ref{alg3} and illustrated graphically in Fig.~\ref{guided_diffusion}.
 
We remark that our goal is to generate realistic samples $\mathbf{z}^{(\text{g})}_{0}$ that, at the same time, are faithful to the original guide $\mathbf{x}^{(\text{g})}$. As analyzed in \cite{meng2021sdedit}, the truncation step $\tau_0$ plays a critical role, as its chosen value directly relates to a trade-off between realism and faithfulness. As can be seen in Fig.~\ref{tau0}, selecting a value of $\tau_0$ closer to 0 would generate a result very close to the guide, but would not produce such a realistic output as an example sampled unconditionally. We must tolerate deviations from the guide to obtain realistic results. However, as $\tau_0$ gets closer to 1, the deviation increases to a point where the conditioning is useless. 



In Fig.~\ref{deviations_tau0}, we analyze the variations from the guide in the generated samples when using different values of $\tau_0$.
For Fig.~\ref{stat_time}, we computed the temporal envelope of the guide represented in Fig.~\ref{tau0} (left) using the root-mean-square procedure and a window of 25\,ms. We then calculated the pairwise differences between the guide envelope and the time envelope of 40 refined samples. We show the standard deviations of the differences in the time envelope for four values of $\tau_0$. Fig.~\ref{stat_freq} was produced in a similar manner by comparing the Bark-smoothed spectral magnitudes between the guide and the refined samples. It can be observed that the variations intensify as we increase $\tau_0$. The prominent peak in Fig.~\ref{stat_time} at about 0.17\,s shows how the thump present in the guide is often suppressed when $\tau_0 \geq 0.5$. We also notice a slight bias in the time envelope for larger values of $\tau_0$, which is related to the effect of clicks and other localized disturbances on the envelope estimation, since more of them are generated when $\tau_0$ is higher. Note that these visualizations were made for one guide in particular; the deviations may vary differently depending on the chosen guide.

\begin{algorithm}[t]
\caption{Guided sampling}
\label{alg3}
\begin{algorithmic}
\Require $T$ (num. steps), $\tau_0$ (truncation step), $\mathbf{x}_\text{g}$ (guide)
\State Sample $\boldsymbol\epsilon \sim \mathcal{N}(\mathbf{0},\mathbf{I})$
\State $\mathbf{z}^{(\text{g})}_{\tau_0} \gets  \alpha_{\tau_0} \mathbf{x}_\text{g} + \sigma_{\tau_0} \boldsymbol\epsilon$
\For{$i=\tau_0 T-1, \dots, 0$}
\State $s \gets \frac{i}{T}$, $\tau \gets  \frac{i+1}{T}$
\State $\mathbf{z}^{(\text{g})}_{s} \gets f(\tau,s) \mathbf{z}^{(\text{g})}_{\tau} - g(\tau,s)\hat{\boldsymbol\epsilon}_{\theta}(\mathbf{z}^{(\text{g})}_{\tau}, \sigma_{\tau})  $
\If{$s>0$} 
\State Sample $\boldsymbol\epsilon \sim \mathcal{N}(\mathbf{0},\mathbf{I})$
\State $\mathbf{z}^{(\text{g})}_{s} \gets \mathbf{z}^{(\text{g})}_{s}+ h(\tau,s)\boldsymbol\epsilon$
\EndIf
\EndFor
\State \Return $\mathbf{z}_0$
\end{algorithmic}
\end{algorithm}

\subsection{Pre-synthesis of the guides}\label{sec:guides}

The concept of guide signals presented in the previous section consists of a draft of the noise produced in a single revolution of a gramophone disk. The guide does not have to be realistic a priori, as it is meant to be refined by the diffusion model. In our experiments, we generate the guides by incorporating prior knowledge on the noise characteristics, applying the findings from \cite{vlimki2008digital}. 

Different timbres of hisses can be generated by filtering white noise with a parametric equalizer. 
The background noise can be filtered uniformly or have variations within a revolution to enforce a more dynamic result.
To simulate deep scratches, we incorporate low-frequency pulses using the model from Eq.~\eqref{thumpmodel}. In addition, we can incorporate rumble sounds by adding extra low-pass  filtered noise. A limitation of this approach is that the generation of softer clicks and crackle is often uncontrollable, due to their usually low energy. Thus, we rely on the diffusion model to synthesize these sounds in an unconditional manner, but assuming that the result will depend on the other characteristics of the guide. Fig.~\ref{fig:guided} shows several representative examples using different types of guide.

\subsection{Generating variations between revolutions}

As mentioned above, gramophone noises are known to have a periodic structure defined by the revolution period, but they often present some noticeable variations between subsequent revolutions. With the methodology described above, we can synthesize realistic noise texture frames with a certain degree of control. However, the method does not take into account the variations between revolutions. In practice, repeating the same generated frame over and over would produce the desired periodic pattern, but the lack of inter-frame variations would make it sound unrealistic to a listener. On the other hand, combining several frames generated independently of each other would solve the inter-frame variation problem, but the low correlation between them would make us miss the characteristic periodic textures of gramophone disks.

Our solution is based on making use of the stochasticity of the sampling algorithm to generate separate outputs corresponding to $N$ different revolutions. This is achieved by bifurcating the reverse diffusion process at a diffusion step $\tau_\text{p}$. When we reach $\tau=\tau_\text{p}$ during sampling, we can generate $N$ different instances of the previous step $s=\tau_\text{p}-\frac{1}{T}$ by applying:
\begin{equation}
    \mathbf{z}_{s,n}=f(\tau,s)\mathbf{z}_{\tau_\text{p}}-g(\tau,s) \hat{\boldsymbol\epsilon}_{\theta}(\mathbf{z}_{\tau}, \sigma_{\tau}) + h(\tau,s)\boldsymbol\epsilon, 
\end{equation}
for each $n$ of the $N$ desired revolutions. Since the noise $\boldsymbol\epsilon$ is purely stochastic, each of the new instances $\mathbf{z}_{s,n}$ will be slightly different from each other. Then, we can continue the reverse diffusion process for each $\mathbf{z}_{s,n}$ independently, following Alg. \ref{alg:sampling} until we obtain the final results $\mathbf{z}_{0,n}$. This idea is also illustrated in Fig.~\ref{guided_diffusion}.

The set of generated noise revolutions can then be recombined in random order by concatenating them to produce endless sounds. Due to the natural noisiness of these sounds, it is not necessary to account for phase mismatches between consecutive revolutions, as they are not usually perceived. Note that, if it was necessary, this problem could be relieved by applying a small amount of overlap.

\begin{figure}[t]
    \centering
         \begin{subfigure}{\linewidth}
    \includegraphics[scale=0.35]{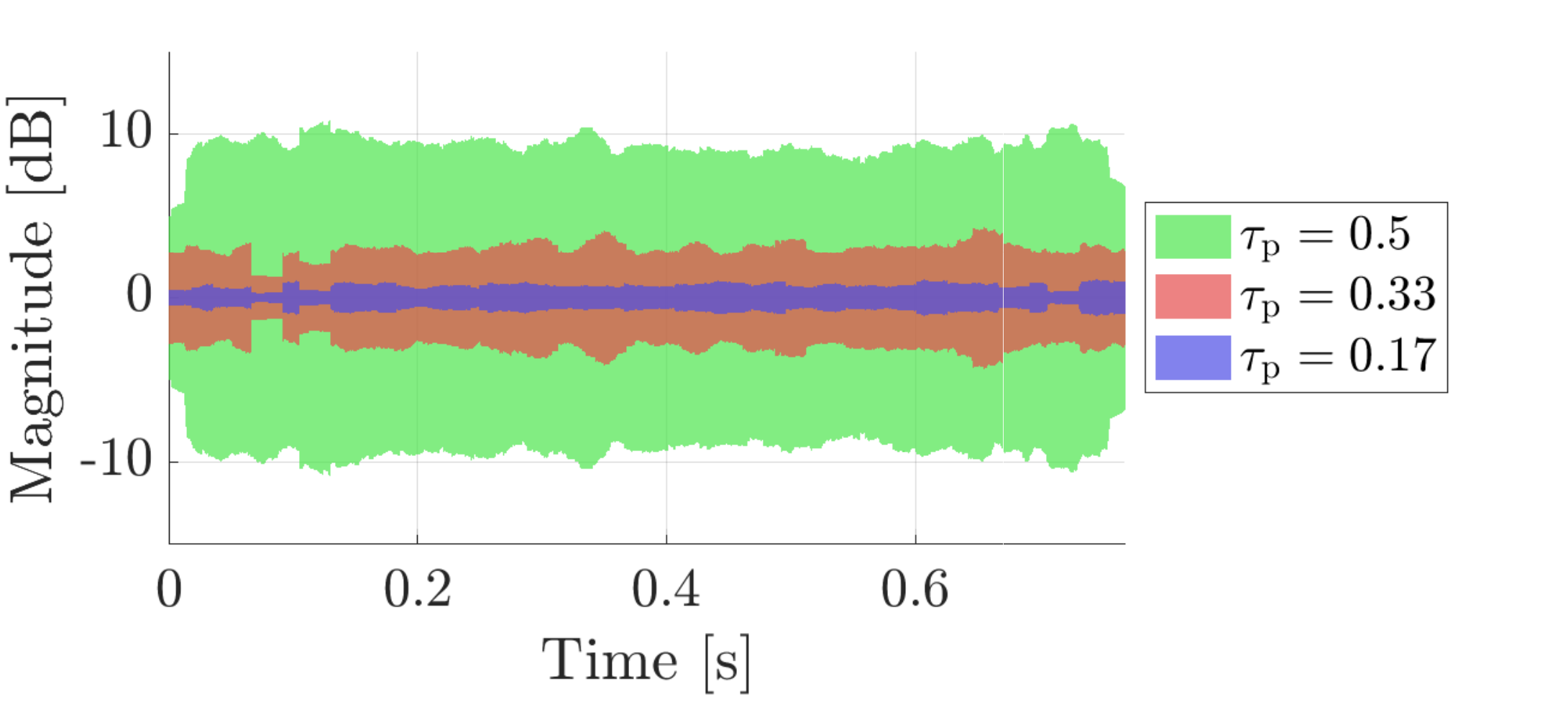}
    \caption{Variations in the temporal envelope for different values of $\tau_\text{p}$.}
    \label{fig:my_label}
        \end{subfigure}
     \begin{subfigure}{\linewidth}
    \includegraphics[scale=0.36]{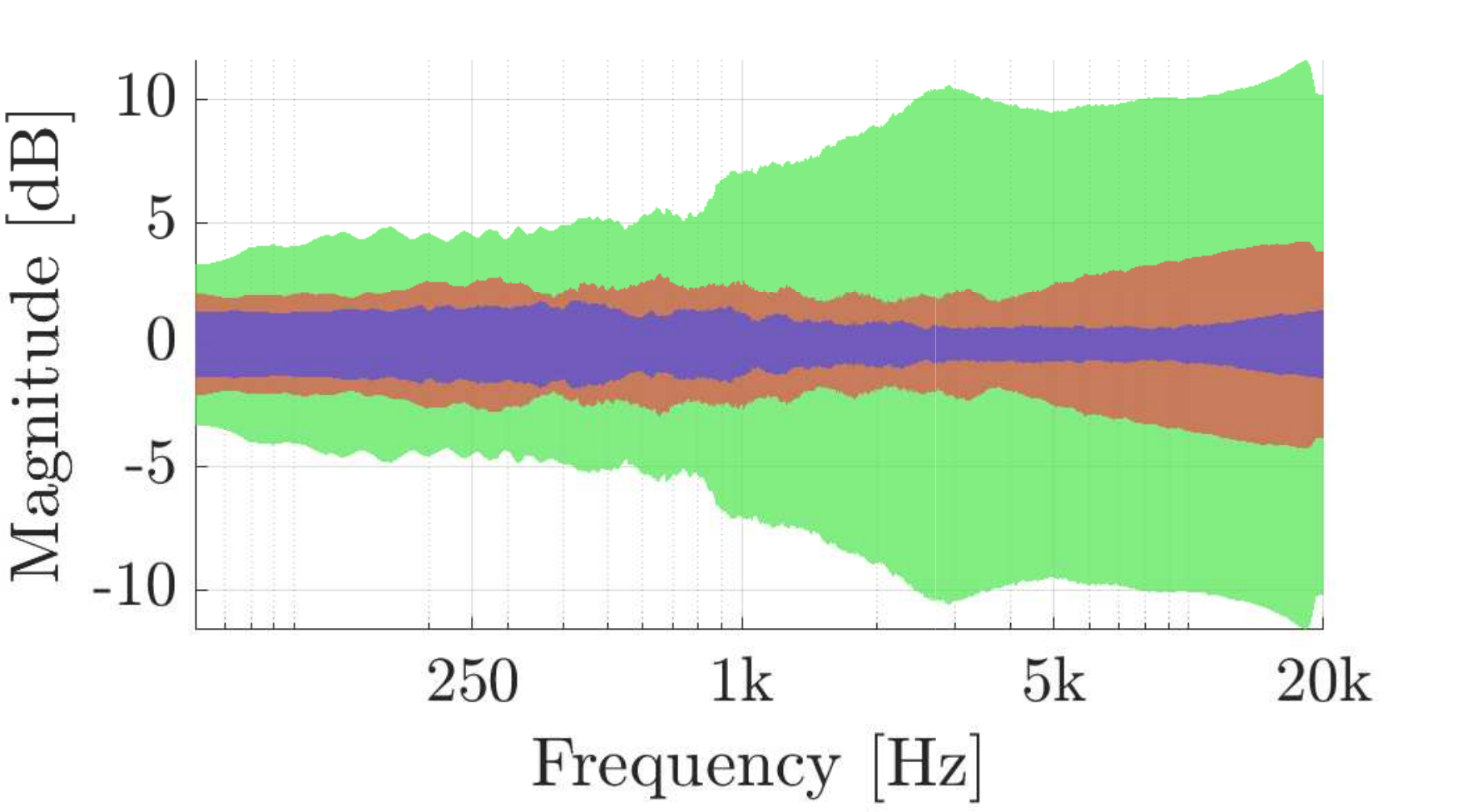}
    \caption{Variations in the spectral envelope for different values of $\tau_\text{p}$.}
         \label{fig:three sin x}
     \end{subfigure}
     \caption{{\it Analysis of the effect of the bifurcation step $\tau_\text{p}$ on the variations between the generated samples representing different disk revolutions.  The standard deviations of the pairwise differences between (a) the temporal envelopes and (b) the Bark-smoothed spectral envelopes are plotted with different colors representing different values of $\tau_\text{p}$.}} 
     \label{revolution_variations}
\end{figure}

As previously stated, our aim is to generate sounds that contain enough variations to seem realistic but still maintain a coherent periodic structure. Similarly to $\tau_0$, the bifurcation step $\tau_\text{p}$ also plays a critical role in the realism of the results. The higher $\tau_\text{p}$ is, the more variations will appear between the generated periods. In Fig.~\ref{revolution_variations}, we analyze the variations in the temporal and spectral envelopes for different values of $\tau_\text{p}$. For both cases, we analyze a single unconditional example, but the sampling algorithm is bifurcated at different diffusion steps $\tau_\text{p}$, obtaining 40 different revolutions. The analysis is carried out by computing the pairwise differences between all the combinations of the 40 output samples. The standard deviations of the pairwise differences are represented in Fig.~\ref{revolution_variations}. As expected, the deviations become wider as we increase $\tau_\text{p}$. We notice that the variations around the temporal envelope follow a uniform pattern, whereas the variations in the spectral envelope are frequency-dependent. We empirically chose the value of $\tau_\text{p}=0.33$, as it often yielded some noticeable variations between revolutions, but the resulting noise periods shared some characteristic similarities.

\section{Subjective Evaluation}\label{sec:results}

\begin{figure}[t]
    \centering
    \includegraphics[width=\columnwidth]{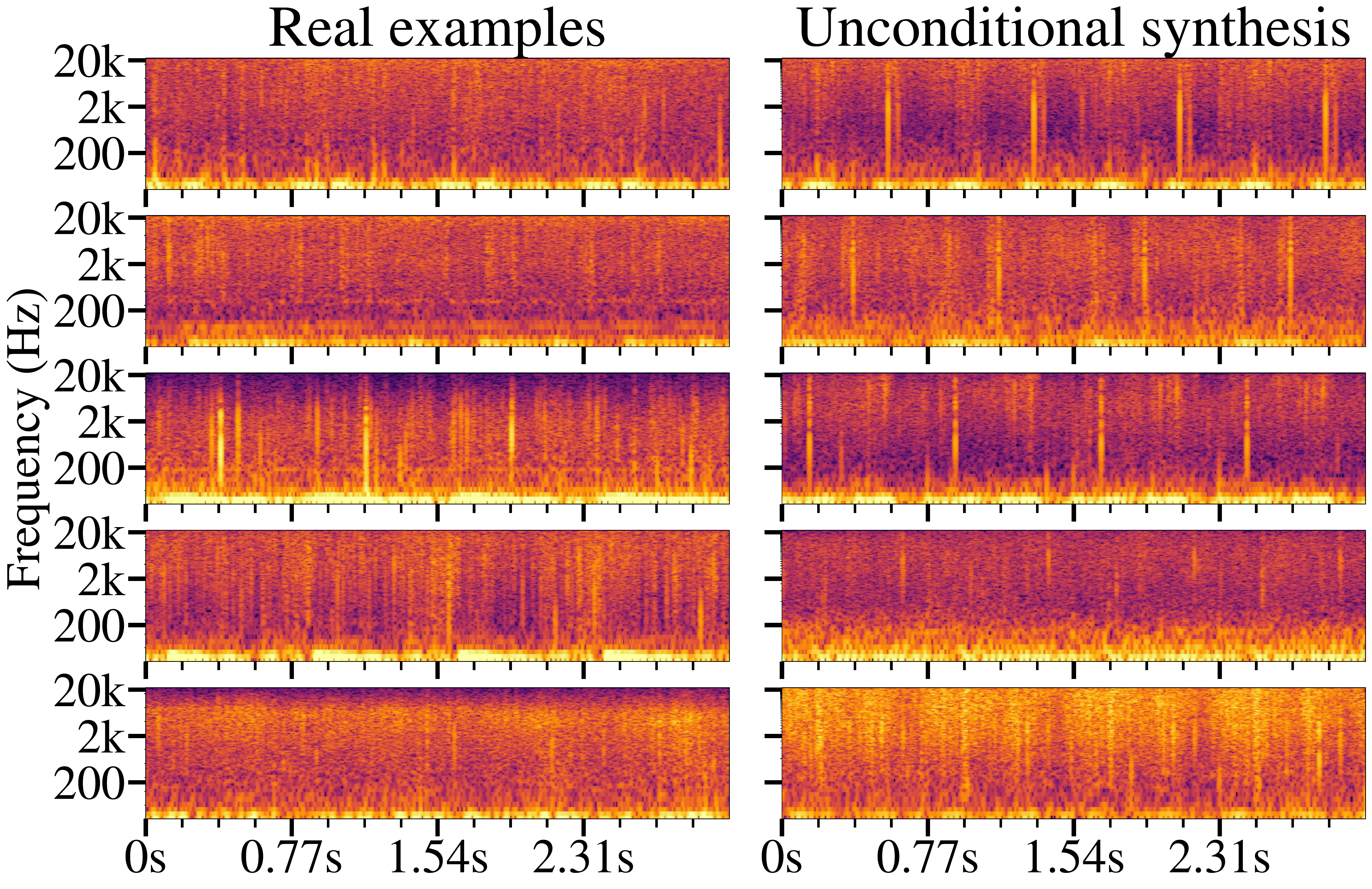}
    \caption{{\it Log-spectrogram representations of a set of real gramophone noise examples and a set of unconditionally synthesized noises using $T=150$ diffusion steps. The length of the synthesized examples correspond to 4 disk revolutions of $L=0.77$\;s; different revolutions are generated by bifurcating the reverse diffusion process at the diffusion step $\tau_\text{p}=0.33$.  }}
    \label{fig:unconditional}
\end{figure}
\begin{figure}[t]
    \centering
    \includegraphics[width=\columnwidth]{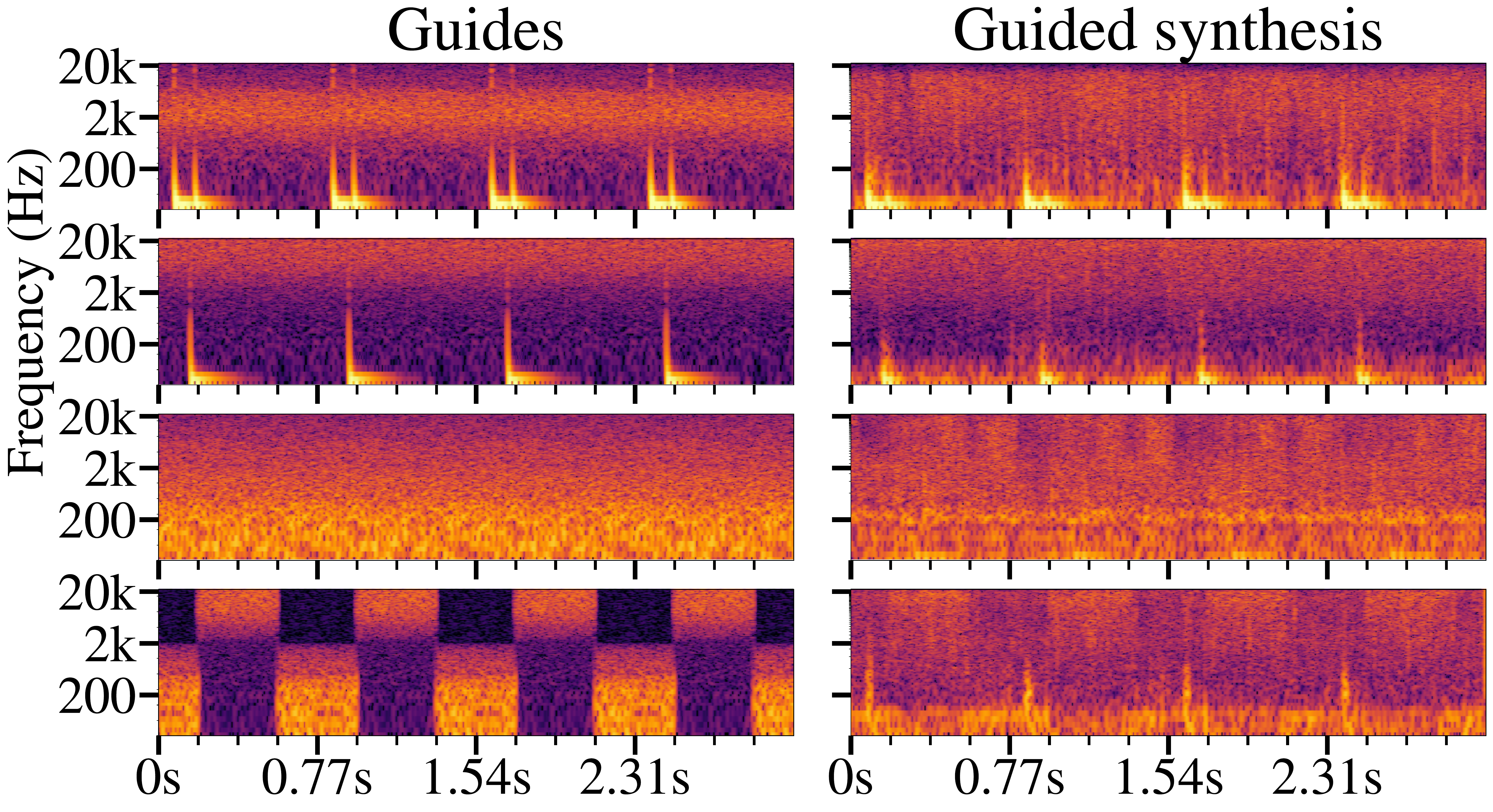}
    \caption{{\it Log-spectrogram representations of different guides and their respective guided synthesis results using $T=150$, $\tau_0=0.5$ and $\tau_\text{p}=0.33$.}}
    \label{fig:guided}
\end{figure}

As represented in Fig.~\ref{fig:unconditional}, the trained diffusion model is capable of unconditionally synthesizing a diverse range of noises with characteristics similar to those of real gramophone recordings. Several results of the use of guided synthesis are also represented in Fig.~\ref{fig:guided}, showing how the generated samples maintain the main characteristics of the guide while generating new and more complex textures. In this section, we investigate whether the generated noise sounds realistic or not.

\begin{table}[t]
\caption{{\it Results of the listening test. Methods that the listeners could not perceive as artificial are highlighted.}}
\label{tab:results}
\centering
\begin{tabular}{@{}llrr@{}}
\toprule
        Synthesis method         & Prob. of success     & p-value\\ \midrule
Digital Audio Antiquing \cite{vlimki2008digital}      & 84.2\% $\pm$ 2.7\%    & <1e-3                 \\
\textbf{Unconditional ($T$=25)}  & \textbf{57.3\% $\pm$ 3.7\%}    & \textbf{0.029}                 \\
\textbf{Unconditional ($T$=150)} & \textbf{52.2\% $\pm$  3.7\%}    & \textbf{0.298}                 \\
Guided  ($T$=150, $\tau_0$=0.33)      & 85.9\%  $\pm$ 2.7\%    & <1e-3                 \\
Guided ($T$=150, $\tau_0$=0.5)        & 70.2\% $\pm$ 3.4\%    & <1e-3                 \\
Guided ($T$=150, $\tau_0$=0.66)        & 60.6\% $\pm$ 3.6\%    & 0.002                 \\\midrule
Random guessing         & 50\% $\pm$ 3.7\% &                       \\ \bottomrule
\end{tabular}
\end{table}

The synthesis of gramophone recording noise is a very particular task, and as far as we are aware, there is no reliable objective metric to evaluate the realism of such methods. As a consequence, we opt to evaluate the realism of the generated noise textures with a subjective listening experiment. The experiment was designed as an AB listening test where the test participants were presented with pairs of noise segments. One of the noises was always a real gramophone extract, and the other was synthesized. For each of the pages, the listeners had to judge or guess which noise segment was the real one. Before taking the test, the participants completed a training session in which the correct answers were revealed and they could learn to distinguish the characteristics of real gramophone recordings. 

The test was divided in two consecutive sessions containing the same pairs of examples but in a different random order. In each session, we included eight examples from six different conditions which were paired with a random real example. The first test condition was included as a baseline containing examples of digital audio antiquing techniques to generate gramophone noise using traditional signal processing methods \cite{vlimki2008digital}, as explained in Sec.~\ref{sec:noises}. We included two unconditional synthesis conditions with a different number of steps ($T$=25 and $T$=150) and three guided synthesis conditions with $T$=150 and different values of the truncation step ($\tau_0$=0.33, $\tau_0$=0.5 and $\tau_0$=0.66). For the guided synthesis examples, we used the same eight guides for every $\tau_0$ value, which were generated as a sum of filtered noise and synthetic thumps, modeled as detailed in Sec.~\ref{sec:guides}.  All the examples are publicly available on the companion web page\footnote{\href{http://research.spa.aalto.fi/publications/papers/dafx22-gramophone-synth/}{research.spa.aalto.fi/publications/papers/dafx22-gramophone-synth/}}.
An extra condition containing static white noise was included four times per session as a control item to detect unreliable participants. 

All in all, the test included 104 signals for each test subject. The test was carried out in an isolated listening booth and took, on average, 30 min to complete. Altogether, 12 volunteers participated in the experiment, although one had to be discarded as they did not identify all the control items. All participants had normal hearing and had previous experience with formal listening tests.

In Table \ref{tab:results}, we report the mean and standard deviations of the probability of success, estimated with a Beta distribution. With probability of success, we refer to the probability of an individual to identify the real gramophone noise when it is compared with a synthesized one. Thus, the lower the probability, the more realistic the synthetic noises are expected to be. For comparison, we include in Table \ref{tab:results} the probability of success when an individual uses a random guessing strategy (e.g. flipping a coin). Note that the reported standard deviations depend on the sample size per condition. We conduct binomial tests to evaluate whether the test answers for each condition are different from random guessing. The resulting p-values are also reported in Table \ref{tab:results}.

The results show how the unconditional examples were practically indistinguishable from the real ones, especially when $T=150$. These results contrast with the participants' performance with the Digital Audio Antiquing examples, as they were able to distinguish them in the majority of the cases. As can be seen, the decrease in the number of diffusion steps does not cause a strong decline in unconditional synthesis performance, indicating that using a large number of steps might not be a critical factor for this specific task. As expected, for the guided synthesis examples, the probability of success increases as $\tau_0$ decreases. In these cases, the test subjects managed to perform better than random guessing, although they were fooled in a significant number of trials. The participants were able to leave comments in the test pages, sharing the strategies they utilized. Some participants focused on the periodicity of the examples, judging too periodic examples to be unreal. Others paid special attention to localizing any kind of music residuals or spotting aliasing artifacts. However, at the end of the experiment, all participants reported that the test was difficult and admitted that they were unsure of their decisions in most cases.



\section{Conclusions}\label{sec:conclusion}

This paper presented a new method to generate realistic gramophone noises. The proposed method is based on a diffusion probabilistic model that generates samples of gramophone noise by progressively denoising Gaussian noise. As demonstrated in a subjective evaluation, the diffusion model is capable of synthesizing highly realistic noises in an unconditional manner.

A method to guide the diffusion model and gain control over the synthesized sound was also proposed. 
However, we notice that this conditioning approach is still not easy to control and often cannot yield highly realistic results, unless the faithfulness to the guide is sacrificed. We leave the study of different conditioning methods for future work.

A known limitation of diffusion models is the high computational cost that they present during the inference stage. 
As a consequence, the presented method is not suitable to work in real-time. 
The reduction of the computational cost of diffusion models is an active field of research. Recent advances, such as progressive distillation \cite{salimans2022progressive}, could help improve the speed, while being orthogonal to most of the methods presented in this work.
Meanwhile, we envision the proposed model being used as a non-real-time effect, with which a few revolutions of gramophone noise could be synthesized and stored in memory. Then, the synthesized noise revolutions could be recombined in random order to generate an ``endless'' background audio track.

Apart from the gramophone noise synthesis task studied in this paper, applying light modifications, the presented model has potential to be applied to the synthesis of other audio textures \cite{caracalla2019sound}. We have experimented applying the presented model to synthesize ``rain'' and ``applause'' sounds. Audio examples of synthetic texture sounds are available in the companion web page.

\section{Acknowledgments}
This work belongs to the activities of the Nordic SMC Network (NordForsk project no.~86892). Special thanks go to Alec Wright for proofreading the manuscript. The authors acknowledge the computational resources provided by the Aalto Science-IT project.

\bibliographystyle{IEEEbib}
\bibliography{references} 

\end{document}

%
%
%